\documentclass[sigconf]{acmart}
\usepackage{algorithm}
\usepackage{algorithmic}
\usepackage{pdfpages}
\usepackage{multirow}
\usepackage[export]{adjustbox} 
\usepackage{bm}
\usepackage{amsmath}
\usepackage{array}
\usepackage{booktabs}
\usepackage{footmisc}

\usepackage{amsthm}
\usepackage{graphicx}
\usepackage{enumitem}

\theoremstyle{definition}

\usepackage{subfigure}
\usepackage{hyperref}
\usepackage{url}
\usepackage{color}
\usepackage{colortbl}
\usepackage{xspace}
\usepackage{tabularx}

\newcommand{\model}{AliBoost\xspace}

\usepackage{amsmath,amsfonts,bm}

\def\eqref#1{equation~\ref{#1}}

\def\1{\bm{1}}

\DeclareMathAlphabet{\mathsfit}{\encodingdefault}{\sfdefault}{m}{sl}
\SetMathAlphabet{\mathsfit}{bold}{\encodingdefault}{\sfdefault}{bx}{n}

\AtBeginDocument{%
  }

\copyrightyear{2025}
\acmYear{2025}
\setcopyright{cc}
\setcctype{by}
\acmConference[KDD '25] {Proceedings of the 31st ACM SIGKDD Conference on Knowledge Discovery and Data Mining V.2}{August 3--7, 2025}{Toronto, ON, Canada.}
\acmBooktitle{Proceedings of the 31st ACM SIGKDD Conference on Knowledge Discovery and Data Mining V.2 (KDD '25), August 3--7, 2025, Toronto, ON, Canada}
\acmISBN{979-8-4007-1454-2/25/08}
\acmDOI{10.1145/3711896.3737188}

\begin{document}

\title{AliBoost: Ecological Boosting Framework in Alibaba Platform}


\author{Qijie Shen}
\authornote{Both authors contributed equally to this paper.}
\affiliation{%
  \institution{Alibaba Group}
  \city{Hangzhou}
  \country{China}
}
\email{qijie.sqj@alibaba-inc.com}

\author{Yuanchen Bei}
\authornotemark[1]
\affiliation{%
  \institution{Zhejiang University}
  \city{Hangzhou}
  \country{China}
}
\email{yuanchenbei@zju.edu.cn}

\author{Zihong Huang}
\affiliation{%
  \institution{Alibaba Group}
  \city{Hangzhou}
  \country{China}
}
\email{huangzihong.hzh@alibaba-inc.com}

\author{Jialin Zhu}
\affiliation{%
  \institution{Alibaba Group}
  \city{Hangzhou}
  \country{China}
}
\email{xiafei.zjl@alibaba-inc.com}

\author{Keqin Xu}
\affiliation{%
  \institution{Alibaba Group}
  \city{Hangzhou}
  \country{China}
}
\email{xukeqin.xkq@alibaba-inc.com}

\author{Boya Du}
\affiliation{%
  \institution{Alibaba Group}
  \city{Hangzhou}
  \country{China}
}
\email{boya.dby@alibaba-inc.com}

\author{Jiawei Tang}
\affiliation{%
  \institution{Alibaba Group}
  \city{Hangzhou}
  \country{China}
}
\email{qingshi.tjw@alibaba-inc.com}

\author{Yuning Jiang}
\affiliation{%
  \institution{Alibaba Group}
  \city{Hangzhou}
  \country{China}
}
\email{mengzhu.jyn@alibaba-inc.com}

\author{Feiran Huang}
\affiliation{
  \institution{Jinan University}
  \city{Guangzhou}
  \country{China}
}
\email{huangfr@jnu.edu.cn}

\author{Xiao Huang}
\affiliation{
  \institution{The Hong Kong Polytechnic University}
  \city{Hong Kong SAR}
  \country{China}
}
\email{xiaohuang@comp.polyu.edu.hk}

\author{Hao Chen}
\authornote{Corresponding author.}
\affiliation{%
  \institution{City University of Macau}
  \city{Macao SAR}
  \country{China}}
\email{sundaychenhao@gmail.com}

\renewcommand{\shortauthors}{Qijie Shen et al.}

\begin{abstract}
Maintaining a healthy ecosystem in billion-scale online platforms is challenging, as users naturally gravitate toward popular items, leaving cold and less-explored items behind. This ``rich-get-richer'' phenomenon hinders the growth of potentially valuable cold items and harms the platform's ecosystem. Existing cold-start models primarily focus on improving initial recommendation performance for cold items but fail to address users' natural preference for popular content. In this paper, we introduce \textbf{AliBoost}, Alibaba's ecological boosting framework, designed to complement user-oriented natural recommendations and foster a healthier ecosystem. AliBoost incorporates a \textit{tiered boosting structure} and \textit{boosting principles} to ensure high-potential items quickly gain exposure while minimizing disruption to low-potential items. To achieve this, we propose the \textit{Stacking Fine-Tuning Cold Predictor} to enhance the foundation CTR model's performance on cold items for accurate CTR and potential prediction. AliBoost then employs an \textit{Item-oriented Bidding Boosting} mechanism to deliver cold items to the most suitable users while balancing boosting speed with user-personalized preferences. Over the past six months, AliBoost has been deployed across Alibaba's mainstream platforms, successfully cold-starting over a billion new items and increasing both clicks and GMV of cold items by over 60\% within 180 days. Extensive online analysis and A/B testing demonstrate the effectiveness of AliBoost in addressing ecological challenges, offering new insights into the design of billion-scale recommender systems.
\end{abstract}

\ccsdesc[500]{Information systems~Recommender systems}
\ccsdesc[500]{Information systems~Business intelligence}
\ccsdesc[300]{Information systems~Online advertising}

\keywords{ecological boosting, item cold-start application, billion-scale recommender systems}



\maketitle

\section{Introduction}

Boosting the ecology of a billion-scale recommendation platform is a long-standing challenge~\cite{chen2024macro,gao2023survey,zhang2024linear,zhao2022understanding}. In natural user-oriented recommendations, users tend to click and purchase more popular items, even when two items have the same content and descriptions~\cite{zhang2025cold,li2021user,he2020lightgcn}. This ``rich-get-richer'' phenomenon hinders the growth of cold items, even if they have the potential to become popular~\cite{fabbri2022exposure,germano2019few,wen2022distributionally}. As cold item stagnation becomes prevalent, item/content producers may become less enthusiastic about uploading new items, leading to a loss of platform vitality. Thus, it is crucial to design a powerful boosting framework to provide more exposure opportunities for new items to foster a healthier ecosystem.

Existing approaches to address this issue, such as cold-start models~\cite{huang2023aldi,huang2024large,liu2023ucc,wei2021contrastive} and debiasing models~\cite{zhou2021contrastive,chen2023bias,yang2023debiased,wang2021deconfounded}, focus on improving the initial performance of new items or compensating for less popular items. For instance, cold-start models like CLCRec~\cite{wei2021contrastive} and ColdLLM~\cite{huang2024large} employ contrastive learning and large language model simulators to optimize cold item embeddings or generate user sequences for cold items. Debiasing models compensate for less popular (long-tail) items by assigning them higher weights. 
DecRS~\cite{wang2021deconfounded} and MACR~\cite{wei2021model} adopt causal graph learning to understand user-item behaviors and alleviate the popularity bias.

\begin{figure}[t]
\centering
    \includegraphics[width=\linewidth]{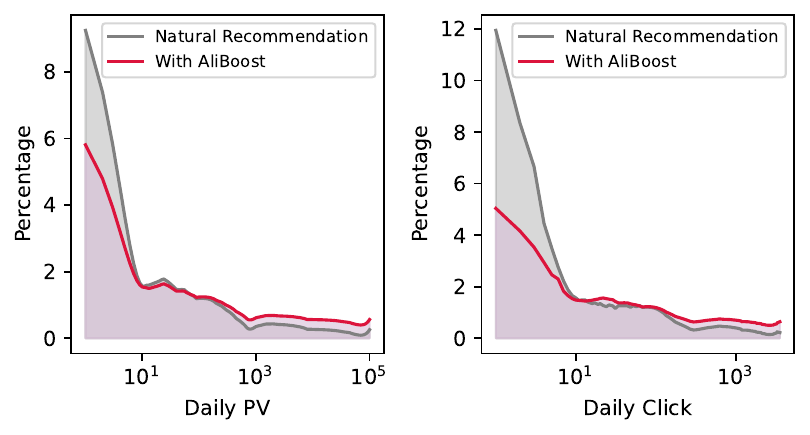}
    \caption{Distribution comparison of daily page views (PV) and daily clicks for cold (new) items after 30 days: natural recommendation vs. \model recommendation.}
\label{fig:intro}
\end{figure}

As shown in~\autoref{fig:intro}, despite the use of cold-start models and debiasing techniques, 41.1\% of cold (new) items fail to achieve 10 daily exposures even 30 days after their launch. This severe ecological problem arises from three primary reasons:
\begin{enumerate}[leftmargin=*]
\item \textbf{Warm Preference Bias}: Users exhibit a natural tendency to click on or interact with warm, popular items, which causes user-oriented recommendation systems to be less inclined to recommend cold items.
\item \textbf{Insufficient Cold-Start Support}: These items often do not receive adequate initial exposure, making it difficult for them to grow and achieve consistent exposure, even after 30 days.
\item \textbf{Lack of Incentivized Exposure}: Even when a new item performs well during the cold-start period, user-oriented recommendation systems often lack mechanisms to ensure sustained exposure over time.
\end{enumerate}

Motivated by these limitations, designing an item-oriented ecological boosting framework to identify and cultivate the potential of cold items, ultimately transforming them into popular items, is a promising avenue. However, there are three key challenges in the design of an item-oriented ecological boosting recommendation:
\begin{enumerate}[leftmargin=*]
\item \textbf{Automatically Boosting Strategy}: Designing effective boosting principles and structures to automatically identify high-potential items and guide them to popularity is a complex task.
\item \textbf{Accurate CTR Prediction}: Cold items have limited behavioral data, making it difficult for CTR models trained on all items to perform well for cold items.
\item \textbf{Optimized Boosting}: Given a fixed budget, designing a suitable boosting framework to promote items to the most relevant users, rather than to uninterested users, is challenging.
\end{enumerate}

To address these challenges, we propose \textbf{AliBoost}, an item-oriented boosting framework for the Alibaba platform, designed to significantly alleviate ecological challenges. As shown in~\autoref{fig:intro}, the percentage of items with fewer than 10 daily PVs decreased from 41.1\% to 24.5\% (a 40\% reduction). Specifically, AliBoost introduces two core principles of boosting recommendations and proposes a complete boosting structure that includes stage-based boosting, evaluation metrics, and promotion and exit mechanisms. Furthermore, we introduce the \textit{Stacking Fine-Tuning Cold Predictor}, a prediction model to accurately estimate the click-through rate (CTR) and potential of cold items. To expose new items to users, we propose the \textit{Item-Oriented Bidding Boosting}, which utilizes a bidding mechanism to deliver cold items to suitable users. Finally, AliBoost provides a comprehensive implementation on the Alibaba platform, ensuring both practicality and scalability.

The key contributions of this paper are as follows:
\begin{itemize}[leftmargin=*]
\item \textbf{A New Recommendation Paradigm}: We propose \textbf{AliBoost}, a novel framework to improve the ecosystem of billion-scale recommender platforms, serving as a pioneer to inspire industrial companies in maintaining a healthy ecosystem.
\item \textbf{Boosting Principles and Designs}: We introduce the principles of boosting recommendations and propose detailed settings for online evaluation, promotion, and removal strategies.
\item \textbf{Boosting Recommender Techniques}: We present the \textit{Stacking Fine-Tuning Cold Predictor} and the \textit{Item-Oriented Bidding Boosting} mechanism to ensure AliBoost can expose new items to suitable users in a controllable and optimal manner.
\item \textbf{Online Experiments and Analysis}: We provide detailed online A/B test results and offer analysis and comparisons before and after deploying AliBoost. Notably, AliBoost increased both clicks and GMV of cold items by over 60\% within 180 days.
\end{itemize}

\section{Related Works}
\subsection{Cold-Start Recommendation}
Cold-start recommendation is a critical challenge in modern recommender systems, particularly when new users or items lack sufficient interaction data to generate accurate recommendations~\cite{huang2023aldi,zhang2025cold}. This problem has significant implications, including decreased user engagement~\cite{ciampaglia2018algorithmic,klimashevskaia2023evaluating}, missed opportunities for personalization~\cite{zhu2021popularity,klimashevskaia2024survey}, and potential reinforcement of biases~\cite{zhang2025cold,wei2021model}. Addressing the problem requires innovative strategies that can effectively leverage limited data for effective cold-start recommendations of new items.
On the one hand, several studies employ contrastive learning~\cite{wei2021contrastive,zhou2023contrastive}, meta-learning~\cite{lee2019melu,zhu2021learning}, or knowledge distillation~\cite{huang2023aldi,wang2021privileged} to transfer the information encoded in warm embeddings to randomly initialized cold embeddings, enabling these cold embeddings to rapidly adapt to the recommender system's existing behavior patterns.
On the other hand, some recent works learn the representations of cold items by generating and synthesizing high-quality interactions for them from a behavioral perspective~\cite{liu2023ucc,huang2024large,liu2024fine}.
Existing methods merely generate synthetic representations or behavior patterns directly for new items at the initial stage of their lifecycle in the systems, without considering the evaluation of changes in cold items throughout their entire lifecycle, as well as the assessment of their quality during growth.

\subsection{Debiased Recommendation}
Debiased recommendation is a significant research direction in modern recommender systems, aiming to mitigate biases such as popularity bias and exposure bias that can lead to unfair and inaccurate recommendations~\cite{zhang2025relieving,chen2023bias,yang2023debiased,zhu2021popularitydynamic}. These biases often result in over-representation of popular items and neglect of long-tail items, affecting recommendation diversity and user satisfaction~\cite{liu2020long,chen2019serendipity}.
To address these challenges, recent studies have explored a rich spectrum of debiasing techniques. For example, some works employ contrastive learning to reduce popularity bias for recommendations~\cite{yang2023debiased,zhang2025relieving}.
Further, some existing works~\cite{wang2021deconfounded,wei2021model} adopt causal graph learning to understand and alleviate the popularity bias.
In this paper, we go beyond focusing on the debiasing process. For the first time in billion-scale systems, we explore a novel mechanism for boosting the lifecycle of new items.

\subsection{Bidding Recommendation}
Real-time bidding (RTB) is a market-based mechanism in which heterogeneous bidders, like advertisers, merchants, or content creators, compete for limited exposure opportunities by submitting bids in real time~\cite{ou2023survey}. As recommender systems and online advertising continue to converge, RTB has been applied to news feeds, search ranking, short-video streams, and other content-distribution scenarios, a line of work collectively referred to as bidding recommendation~\cite{zhang2023personalized,ji2024learning,carrion2023blending}. Nevertheless, executing real-time auctions over billions of candidates under strict latency budgets while balancing short-term revenue with long-term ecosystem health remains a central challenge for the systems.
Therefore, in this paper, we investigate how to automatically allocate greater exposure to higher-quality items according to their intrinsic merit, thereby sustaining a healthy industrial-scale recommendation ecosystem.

\section{Methodology}
In this section, we first present the collaboration between boosting recommendations and natural recommendations, along with the design of a Tiered Boosting Structure. Subsequently, we introduce the Stacking Fine-Tuning CTR Predictor, which aims to increase the click-through rate prediction accuracy for new items. We then propose the Item-Oriented Bidding Boosting, demonstrating how to optimally expose cold items within the appointed budget provided by the boosting structure. Finally, we describe the implementation details of \model within the Alibaba platform.

\subsection{Overall Framework}
\begin{figure}[t]
\centering
    \includegraphics[width=\linewidth]{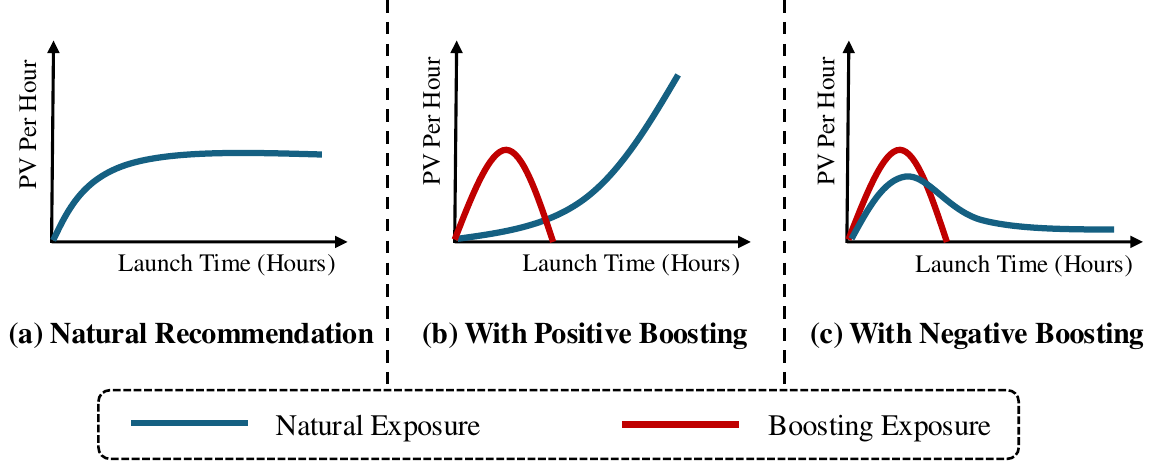}
    \caption{Toy example of three types of recommendation situations.}
\label{fig:boost_illu}
\end{figure}

As illustrated in~\autoref{fig:boost_illu}-(a) (\underline{Natural Recommendation}), relying solely on the natural recommendation system leads to a prolonged cold-start period. To address this, in \model framework, as shown in~\autoref{fig:boost_illu}-(b) (\underline{Positive Boosting}), our boosting recommendation provides new items with high-quality exposure to accelerate the cold-start phase while maintaining strong recommendation performance. This approach enables boosted items to leverage the initial exposure as a foundation to gain additional opportunities through the natural recommendation system. However, as depicted in~\autoref{fig:boost_illu}-(c) (\underline{Negative Boosting}), a suboptimal boosting strategy may result in low-quality exposures during the boosting period, leading to poor recommendation performance. Consequently, such items may fail to secure further exposure through the natural recommendation system.

\begin{figure*}[t]
\centering
    \includegraphics[width=\linewidth]{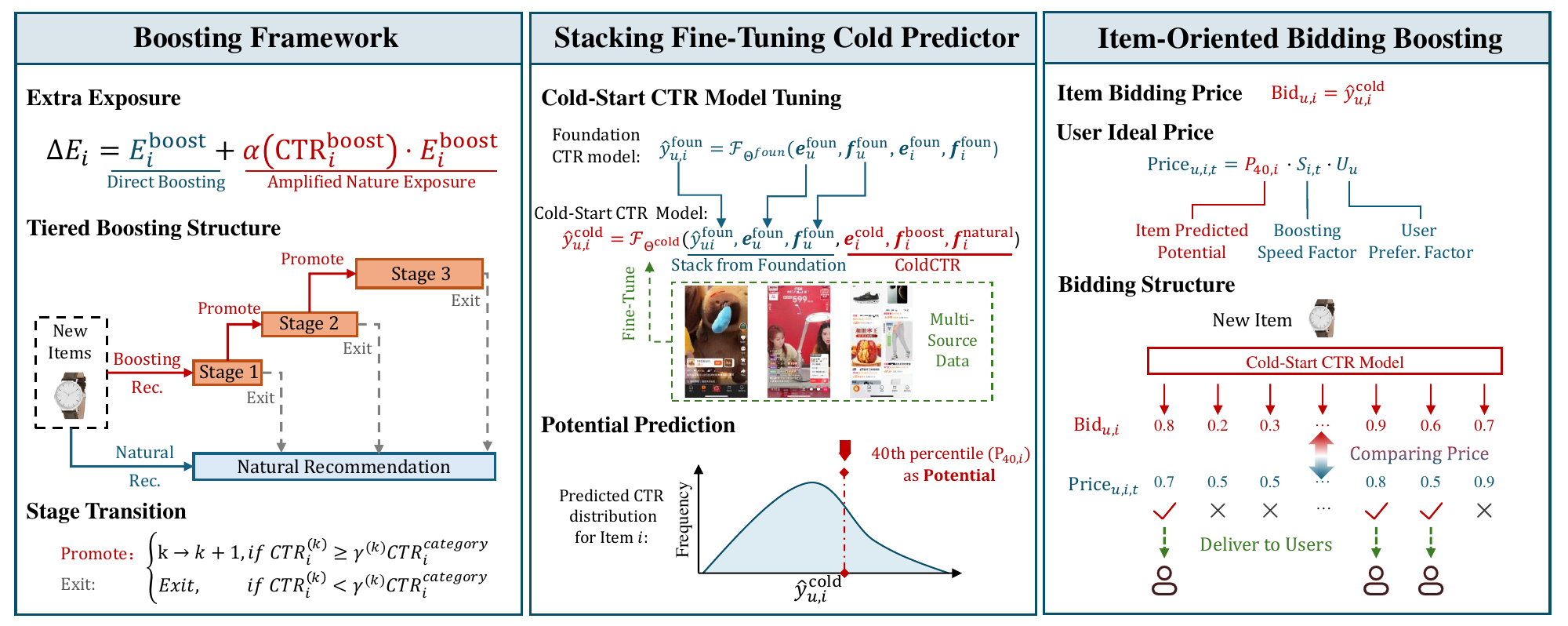}
    \caption{Overall architecture of the \model framework.}
\label{fig:framework}
\end{figure*}

To formalize the interaction between boosting recommendations and natural recommendations, the total incremental exposure $\Delta E_i$ for an item $i$ can be expressed as:
\begin{equation}
\Delta E_i = \underbrace{E^{\text{boost}}_i}_{\text{Direct boosting expo.}} 
+ \underbrace{\alpha(\text{CTR}^{\text{boost}}_i)\\ \cdot E^{\text{boost}}_i}_{\text{Amplified natural expo.}},
\end{equation}
where the first term, $E^{\text{boost}}_i$, represents the direct exposure allocated to item $i$ via the boosting framework. The second term, $\alpha(\text{CTR}^{\text{boost}}_i) \cdot E^{\text{boost}}_i$, captures the amplified effect provided by the natural recommendation system, which is influenced by the recommendation performance achieved during the boosting period. Here, $\text{CTR}^{\text{boost}}_i$ denotes the click-through rate (CTR) of item $i$ during the boosting phase, while the amplification function $\alpha(\text{CTR}^{\text{boost}}_i)$ models the non-linear relationship between boosted CTR and additional organic exposure. Typically, $\alpha(\cdot)$ exhibits exponential growth for items with high $\text{CTR}^{\text{boost}}_i$, amplifying the exposure of high-performing items through the natural recommendation system.

This formulation captures two fundamental phenomena:
\begin{itemize}[leftmargin=*]
\item \textbf{Direct Boosting Effect}: Initial boosting exposure provides essential cold-start exposure opportunities for new items.
\item \textbf{Non-linear Amplification}: High-performance items gain exponentially more natural exposure with the help of the boosting recommendation.
\end{itemize}

Based on this understanding, we propose the following two core boosting principles:

\textbf{Performance-Driven Boosting}:
New items with higher boosting recommendation performance should receive more boosting opportunities for user exposure:
\begin{equation}
E^{\text{boost}}_i \propto \text{CTR}^{\text{boost}}_i.
\end{equation}

\textbf{Non-Disturbance Principle}:
Boosting recommendations should maintain recommendation quality above the platform and category average, and should not disturb the natural recommendation:
\begin{equation}
\mathbb{E}[\text{CTR}^{\text{boost}}_i] \geq \gamma \cdot \text{CTR}^{\text{natural}}_i,
\end{equation}
where $\mathbb{E}[\text{CTR}^{\text{boost}}_i]$ denotes the expectation of the boosting CTR for item $i$, $\text{CTR}^{\text{natural}}_i$ denotes the natural recommendation CTR, and $\gamma$ is a safety factor (typically $\gamma=1.2$).

\subsection{Tiered Boosting Structure}
The proposed boosting framework adopts a tiered structure to systematically allocate exposure opportunities to cold items while dynamically evaluating their performance. This structure ensures a balance between maximizing the potential of promising items and maintaining overall recommendation quality. The tiered structure consists of the following key components:

\paragraph{\textbf{Boosting Stages}}
The boosting pipeline defines a structured progression for cold items through a series of stages, each associated with an incrementally larger exposure budget. This hierarchical design enables the framework to gradually allocate exposure opportunities, ensuring that items demonstrate sufficient potential before advancing to higher exposure levels. Formally, let $B_i^{(k)}$ represent the exposure budget assigned to item $i$ at stage $k$. The exposure budget for each stage satisfies the condition:
\begin{equation}
    B_i^{(1)} < B_i^{(2)} < \cdots < B_i^{(K)},
\end{equation}
where $K$ is the total number of boosting stages. At stage $k$, an item $i$ enjoys the cumulative exposure opportunities from all preceding stages to ensure that items promoted to higher stages benefit from both the exposure allocated at that stage and the residual exposure from earlier stages. Noted, we can directly assign an item $i$ stages according to the $\text{Stage}_i$ in Eq. (\ref{eq:grade}).

\paragraph{\textbf{Promotion and Exit Mechanism}}
To ensure the effective allocation of resources, the boosting framework incorporates a dynamic promotion and phase-out mechanism. At the end of each stage, an item's performance is evaluated against category-specific benchmarks. Specifically, an item $i$ is promoted to the next stage if its boosting performance satisfies:
\begin{equation}
    \text{CTR}_i^{(k)} \geq \gamma^{(k)} \cdot \text{CTR}_i^{\text{category}},
\end{equation}
where $\text{CTR}_{\text{category}}$ represents the average CTR for items within the same category, $\gamma^{(k)} > 1$ is a safety factor ensuring that boosted items exceed the average CTR of the same category. Further, to ensure the quality of the items, the thresholds for the promotions are progressively increased. Therefore, $\gamma^{(1)}< \cdots < \gamma^{(k)} < \cdots < \gamma^{(K)}$.

Conversely, an item exits the boosting process if its performance falls below the threshold:
\begin{equation}
    \text{CTR}_i^{(k)} < \gamma^{(k)} \cdot \text{CTR}_i^{\text{category}}.
\end{equation}
Together, the promotion and exit conditions can be compactly expressed as:
\begin{equation}
    \text{Stage Transition:} \quad 
    \begin{cases} 
    k \to k+1, & \text{if } \text{CTR}_i^{(k)} \geq \gamma^{(k)} \cdot \text{CTR}_i^{\text{category}}, \\
    \text{Exit}, & \text{if } \text{CTR}_i^{(k)} < \gamma^{(k)} \cdot \text{CTR}_i^{\text{category}}.
    \end{cases}
\end{equation}

Based on the promotion and exit mechanism, the total exposure budget $B^{boost}_i$ allocated to an item $i$ is determined as:
\begin{equation}
    B^{\text{boost}}_i = \sum_{k=1}^{K} \mathbb{I}\left[\text{CTR}_i^{(k)} \geq \gamma^{(k)} \cdot \text{CTR}_i^{\text{category}} \right] \cdot B_i^{(k)},
\end{equation}
where $\mathbb{I}[\cdot]$ is an indicator function that ensures budget allocation only for items meeting the promotion criteria, and $K$ is the maximum number of boosting stages.
\subsection{\textbf{Stacking Fine-Tuning Cold Predictor}}

In billion-scale recommendation systems, predicting the click-through rate (CTR) for new items presents unique challenges due to their lack of historical interaction data. While large-scale CTR foundation models are widely used to provide general predictions across the platform, their performance on cold items often falls short due to insufficient representation of item-specific and boosting-related features. To address this issue, we propose a Stacking Fine-Tuning CTR Predictor tailored to new items, which augments the foundation CTR model with enriched features and fine-tuning strategies. Furthermore, we introduce a potential prediction mechanism to dynamically assess and rank the growth potential of new items based on their predicted CTR distributions.

\subsubsection{\textbf{Stacking Structure}}

The cold-start CTR model builds on the outputs of the platform’s foundation CTR model, incorporating additional cold item-specific features to improve accuracy during the boosting phase. Let $\hat{y}_{u,i}^{\text{foun}}$ denote the CTR predicted by the foundation model $f_{\theta^{\text{foun}}}$ for a user $u$ and an item $i$, which can be defined as:
\begin{equation}
    \hat{y}^{\text{foun}}_{u,i}=f_{\theta^{\text{foun}}}(\mathbf{e}_u^{\text{foun}}, \mathbf{f}_u^{\text{foun}}, \mathbf{e}_i^{\text{foun}}, \mathbf{f}_i^{\text{foun}}),
\end{equation}
where $\mathbf{e}_u^{\text{foun}}$ and $\mathbf{e}_i^{\text{foun}}$ are the foundational user/item embeddings, $\mathbf{f}_u^{\text{foun}}$ and $\mathbf{f}_i^{\text{foun}}$ are the basic user/item features for the foundation CTR predictor in the natural recommendation. 
Then, the stacking structure introduces additional features derived from both natural and boosting recommendation statistics. Specifically, the feature vector $\mathbf{x}^{\text{stack}}_{u,i}$ for the stacked model is constructed as:

\begin{equation}
    \mathbf{x}^{\text{stack}}_{u,i} = \left[\underbrace{\hat{y}_{u,i}^{\text{foun}}, \mathbf{e}_u^{\text{foun}}, \mathbf{f}_u^{\text{foun}}}_{\text{Foundation CTR}}, \underbrace{\mathbf{e}_i^{\text{cold}},  \mathbf{f}_i^{\text{boost}}, \mathbf{f}_{i}^{\text{natural}}}_{\text{Cold CTR}} \right],
\end{equation}
where $\mathbf{e}_i^{\text{cold}}$ is the stacking embedding tuned for cold items, $\mathbf{f}_i^{\text{natural}}$ includes real-time feature streams for item $i$, and $\mathbf{f}_i^{\text{boost}}$ includes cold item features such as item attributes, metadata, and contextual signals (e.g., category, upload time).

For the cold item $i$, the stacked model employs a multi-layer perception to learn the final CTR prediction $\hat{y}^{\text{cold}}_{u,i}$:
\begin{equation}
    \hat{y}_{u,i}^{\text{cold}} = \sigma\left(\mathbf{W}_L \phi_L \left( \cdots \phi_1 (\mathbf{W}_1 \mathbf{x}^{\text{stack}}_{u,i} + \mathbf{b}_1) + \cdots \right) + \mathbf{b}_L \right),
\end{equation}
where $\phi_l(\cdot)$ represents the activation function of the $l$-th layer, $\mathbf{W}_l$ and $\mathbf{b}_l$ are the weights and biases, and $\sigma(\cdot)$ is the sigmoid function to constrain the output between 0 and 1.

\subsubsection{\textbf{Cold-Start CTR Model Tuning}}
Fine-tuning the stacked model is essential to adapt the CTR predictions to the specific characteristics of new items. We propose a two-step fine-tuning process: data enrichment and loss optimization.

\paragraph{\textbf{Data Enrichment}}
To ensure sufficient coverage of new items, we augment the training dataset by including enriched samples from diverse data sources, such as products, advertisements, and short videos. Each sample includes both natural recommendation and boosting recommendation data to capture the evolving user-item interaction dynamics during the cold-start period. The enriched dataset is denoted as $\mathcal{S}$, which contains $(u,i,y_{u,i}^{(s)},s)$.

\paragraph{\textbf{Loss Optimization}}
The overall loss function for fine-tuning incorporates weighted contributions from different data sources to balance their impact on the final prediction. Formally, the loss function is defined as:

\begin{equation}
    \mathcal{L} = \sum_{(u,i,y_{u,i}^{(s)},s)\in\mathcal{S}} \omega_{s} \cdot \mathcal{L}_{\text{rec}}(y_{u,i}, \hat{y}_{u,i}) + \alpha \cdot ||\Theta||_2^2,
\end{equation}
\begin{equation}
    \mathcal{L}_{\text{rec}} = -y_{u,i} \log(\hat{y}_{u,i}^{\text{cold}}) - (1 - y_{u,i}) \log(1 - \hat{y}_{u,i}^{\text{cold}}),
\end{equation}
where $y_{u,i}$ as the ground-truth label and $\hat{y}_{u,i}^{\text{cold}}$ as the predicted CTR, $\omega_{s}$ is a weight factor for each source of samples, determined by its data source, and $\Theta$ is the model parameter, and $\alpha$ is the regularization coefficient.

\subsubsection{\textbf{Potential Prediction}}
Assessing the potential of new items is crucial for prioritizing high-growth items during the boosting phase. Using the cold-start fine-tuned CTR model, we predict the CTR distribution for each new item by sampling a large user set $\mathcal{U}_{\text{sample}}$ of size $N$. For each user $u \in \mathcal{U}_{\text{sample}}$, the predicted CTR is calculated as $\hat{y}_{u,i}$, resulting in a CTR distribution:
\begin{equation}
    \mathcal{D}_i = \{\hat{y}_{u,i}^{\text{cold}} : u \in \mathcal{U}_{\text{sample}}\}.
\end{equation}

To quantify the potential of item $i$, we compute the 40th percentile of the predicted CTR distribution $\mathcal{D}_i$, denoted as $P_{40,i}$. This value serves as a robust measure of the item's performance, as it captures the CTR threshold that 40\% of the sampled users are predicted to exceed.

For potential grading, we assign grades to items based on their $P_{40,i}$ values relative to the distribution of $P_{40}$ values across all items. Let $\mathcal{P}_{40} = \{P_{40,i} : i \in \mathcal{I}\}$ be the set of 40th percentile values for all items, where $\mathcal{I}$ is the set of all items. We define the rank of item $i$, denoted as $r_i$, as the percentage of items with $P_{40}$ values lower than or equal to $P_{40,i}$:
\begin{equation}
r_i = \frac{|{j \in \mathcal{I} : P_{40,j} \leq P_{40,i}}|}{|\mathcal{I}|} \times 100
\end{equation}

Based on the rank $r_i$, we assign stages in the tiered boosting structure according to the following criteria:
\begin{equation}
\label{eq:grade}
\text{Stage}_i =
\begin{cases}
1, & \text{if } r_i <70\%, \\
2, & \text{if } 70\% \leq r_i < 90\%, \\
3, & \text{if } r_i \geq 90\%.
\end{cases}
\end{equation}

\subsection{Item-Oriented Bidding Boosting}
The core motivation behind the item-orient bidding is twofold: (1) to ensure smooth and controllable boosting exposure over a specified period (e.g., 3 days) rather than concentrating the exposure within a short timeframe; and (2) to optimize boosting by exposing new items to the most suitable users, rather than uninterested or dissatisfied users.
However, achieving these objectives presents two key challenges:
\begin{enumerate}[label=(\roman*),leftmargin=*]
\item \textbf{Uncontrollable User Behavior}: Users' online and offline behaviors are unpredictable, making it difficult to deliver recommendations as planned.
\item \textbf{Uncontrollable boosting Numbers}: Cold items may be delivered either too frequently (over-boosting) or too infrequently (under-boosting), making it challenging to strike the right balance between exposure and user engagement.
\end{enumerate}

Many existing systems rely on rule-based heuristics, such as using memory databases to track boosting counts \cite{zhang2015memory,konstas2009social,nilashi2018recommender}. However, these approaches can be inefficient, as cold items may be delivered to suboptimal users who come online earlier than more suitable candidates, ultimately hindering the growth and performance optimization of these items.

\subsubsection{\textbf{Bidding Structure}}
To overcome these limitations, we introduce the item-oriented bidding structure. For a given user-item pair, the item has a bidding price, and the user has an ideal price. If the bidding price exceeds the ideal price, the item is delivered to the user; otherwise, the item is not delivered. This relationship can be expressed as:
\begin{equation}
\text{Deliver}_{u, i, t} =
\begin{cases}
1, & \text{if } \text{Bid}_{u,i} > \text{Price}_{u,i,t}, \\
0, & \text{otherwise}.
\end{cases}
\end{equation}

The item's bidding score is determined by the fine-tuned cold-start CTR predictor, denoted as $ \text{Bid}_{u,i} = \hat{y}_{u,i}^{\text{cold}}$. The user's ideal price is based on a basic score and influenced by two factors: the user preference factor $U_{u}$ and the boosting speed factor $S_{i,t}$. This formulation can be given as:
\begin{equation}
\text{Price}_{u,i,t} = P_{40,i} \cdot S_{i,t} \cdot U_{u},
\end{equation}
where the basic score is given by the 40th percentile of the cold-start fine-tuned CTR predictions for item $i$, and the user preference factor and boosting speed factor vary for each time slot $t \in \mathcal{T}$.

\subsubsection{\textbf{Boosting Speed Factor}}
The boosting speed factor $S_{i,t}$ controls the budget consumption rate for item $i$ at time slot $t$. The target boosting speed is denoted as $V_{i,t}^{\text{target}}$, while the actual boosting speed is denoted as $V_{i,t}$. The speed error $E_{i,t}$ is defined as:
\begin{equation}
E_{i,t} = V_{i,t}/V_{i,t}^{\text{target}}.
\end{equation}
The initial boosting speed factor is derived from the speed error:
\begin{equation}
S_{i,t} = E_{i,t}.
\end{equation}
Further, this approach ensures that the boosting rate is adjusted dynamically:
\begin{itemize} [leftmargin=*]
\item If $V_{i,t} < V_{i,t}^{\text{(target)}}$, $S_{i,t}$ decreases, lowering the ideal price and increasing the boosting speed $V_{i,t+1}$.
\item If $V_{i,t} > V_{i,t}^{\text{(target)}}$, $S_{i,t}$ increases, raising the ideal price and decreasing the boosting speed $V_{i,t+1}$.
\end{itemize}

To prevent large fluctuations, we incorporate the speed errors from previous time slots to maintain stability:
\begin{equation}
S_{i,t} = \delta_p \cdot E_{i,t} + \delta_q \cdot E_{i,t-1} + \delta_d \cdot E_{i,t-2},
\end{equation}
where $\delta_p$, $\delta_q$, and $\delta_d$ are hyperparameters.

\subsubsection{\textbf{User Preference Factor}}
We consider user preference from two aspects: \textbf{user fatigue} and \textbf{user activity}. First, the user fatigue factor $U_{\text{tired}}$ models the user's tiredness towards new items, defined as the number of consecutive exposures to new products without a click. Second, for user activity, more active users are more likely to explore new items. We evaluate user activity using 10 grades, where the most active users have $U_{\text{active}} = 10$, and the least active users have $U_{\text{active}} = 1$.

Based on these settings, the overall user preference factor can be defined as:
\begin{equation}
U_u = \ln(U_{\text{tired}}) \cdot (U_{\text{active}})^{-\frac{1}{2}}.
\end{equation}

This formulation ensures that user fatigue and activity levels are considered when determining the ideal price for each user, enabling the item-oriented bidding to optimize the exposure of new items to the most receptive users while maintaining a smooth and controllable boosting process.

Furthermore, we provide a detailed description of the industrial deployment of \model in the Alibaba recommendation platform in Appendix~\ref{sec:online}.

\section{Experiments}
In this section, we present the online analysis and experiments conducted on the leading e-commerce platform Alibaba, aiming to address the following research questions:
\textbf{RQ1:} How does \model improve the ecological environment of Alibaba?
\textbf{RQ2:} What is the effect of the Boosting structure components in \model?
\textbf{RQ3:} How does \model compare with foundational recommendation models in both offline and online settings?
\textbf{RQ4:} How does the item-oriented bidding contribute to improving recommendations?
\textbf{RQ5:} How does \model influence the potential Matthew effect within the system?

\subsection{Experimental Setup}

\subsubsection{\textbf{Platform, Datasets, and Online Environment}}

Alibaba is a global leading platform in e-commerce, retail, and technology, offering intelligent solutions across diverse scenarios. \model, a core component of Alibaba's recommendation system, is designed to handle these scenarios, impacting billions of users and items.

We utilize two key datasets in this study: the foundation CTR training dataset and the cold-start fine-tuning dataset. 
The foundation CTR training dataset is designed to train the foundational CTR model. It contains user-item interaction data collected over the past six months, covering 0.4 billion users, 0.3 billion items, 24,568 categories, and 142 billion interactions. This dataset provides a solid foundation for building an effective CTR prediction model.  
The cold-start fine-tuning dataset is used to fine-tune the cold-start CTR model, enabling it to better handle new items. This dataset contains user-item behavior data for items launched in the past six months. Specifically, it spans three critical scenarios: e-commerce, live streaming, and short videos. It includes 0.25 billion users, 1.5 billion items, and 7.1 billion interactions.

All online experiments are conducted on Alibaba's primary online system. The online A/B testing environment supports large-scale experiments, with daily metrics including daily 0.3 billion users, daily 1 million items, and daily 10 billion interactions. The detailed statistics of these datasets are presented in Table~\ref{tab:data}.

\begin{table}[htbp]
  \centering
  \small
  \caption{Statistics of the data (B: billions, M: millions).}
  \resizebox{\linewidth}{!}{
    \begin{tabular}{c|c|c|c|c}
    \toprule
    Data & Users & Items & Catagories & Interactions \\
    \midrule
    Foundation Training  & 0.4B & 0.3B & 24,568 & 142B \\
    Cold-Start Fine-Tuning & 0.25B & 1.5B & 9,567 & 7.1B \\
    
    Online A/B Test & Daily 0.3B & Daily 1M & Full & Daily 10B \\
    \bottomrule
    \end{tabular}%
  }
  \label{tab:data}
\end{table}%

\subsubsection{\textbf{Evaluation Metrics}}
We evaluate our approach using two categories of metrics: effectiveness metrics and growth metrics. These complementary measures enable us to assess both the immediate performance and long-term sustainability of the recommendation system.
\paragraph{(i) Effectiveness Metrics}
To quantify the immediate impact of recommendations, we employ four key performance indicators:
\begin{itemize}[leftmargin=*]
    \item \textbf{Page Views (PV)}: The total number of times recommended items are displayed to users, representing the reach of our system.
    \item \textbf{Click-Through Rate (CTR)}: The ratio of user clicks to page views, calculated as:
    \begin{equation}
        \text{CTR} = \frac{\text{Number of Clicks}}{\text{Page Views}} \times 100\%
    \end{equation}
    \item \textbf{Click}: The number of the user clicking.
    \item \textbf{Pay}: The number of the payments.
    \item \textbf{Gross Merchandise Value (GMV)}: The total monetary value generated through recommended purchases, directly measuring business impact.
\end{itemize}

\paragraph{(ii) Growth Metrics}
To evaluate the long-term sustainability and item discovery capabilities, we monitor:
\begin{itemize}[leftmargin=*]
    \item \textbf{Traffic Share}: The exposure proportion of newly listed items within the organic recommendation pipeline:
    \begin{equation}
        \text{Traffic Share} = \frac{\text{cold item PV}}{\text{Total PV}} \times 100\%
    \end{equation}
    
    \item \textbf{Return on Investment (ROI)}: The efficiency of our boosting strategy for cold items:
    \begin{equation}
        \text{ROI} = \frac{\text{Natural Recommendation PV for cold items}}{\text{Boost PV for cold items}}
    \end{equation}
    
    \item \textbf{Hot Item Count}: The number of items achieving substantial daily exposure (exceeding 10,000 page views), demonstrating the system's ability to identify and promote trending content effectively.
\end{itemize}

\subsubsection{\textbf{Implementation Settings}}
For all models, the embedding size is fixed to 256 for fair recommendations. For the user A/B test, we randomly hashed and divided users into buckets, with specific buckets designated as the control group and others as the experiment group. This setup was designed to evaluate model accuracy improvements, aligning with the standard A/B testing paradigm used in most recommendation systems.
For the item A/B test, we randomly hashed and divided items into buckets, with specific buckets serving as the control group and others as the experiment group. This approach was well-suited to observe the long-term growth performance of items under different mechanisms.

\subsection{Main Results (RQ1)}

\begin{table}[b]
\centering
\caption{Performance improvement of \model in Alibaba. We first compare the overall platform improvements and then analyze the impact of boosting based on the launch time of cold items over different time periods.}
\begin{tabular}{lcccc}
\toprule
\textbf{Metric} & \textbf{PV} & \textbf{CLICK} & \textbf{PAY} & \textbf{GMV} \\ 
\midrule
\textbf{Platform Overall} & +2.01\% & +4.51\% & +3.96\% & +4.69\% \\ \midrule
\textbf{Boosted 3 Days}   & +16.50\% & +8.30\% & +6.50\% & +17.90\% \\ 
\textbf{Boosted 7 Days}   & +29.20\% & +19.60\% & +18.20\% & +25.30\% \\ 
\textbf{Boosted 30 Days}  & +44.93\% & +44.59\% & +40.16\% & +40.59\% \\ 
\textbf{Boosted 60 Days}  & +46.36\% & +55.15\% & +56.24\% & +52.06\% \\ 
\textbf{Boosted 90 Days}  & +51.80\% & +57.65\% & +55.13\% & +58.53\% \\ 
\textbf{Boosted 120 Days} & +58.94\% & +61.68\% & +65.71\% & +67.75\% \\ 
\textbf{Boosted 150 Days} & +63.01\% & +73.47\% & +63.76\% & +69.48\% \\ 
\textbf{Boosted 180 Days} & +65.87\% & +76.09\% & +74.26\% & +72.03\% \\ 
\bottomrule
\end{tabular}
\label{tab:overall_ab}
\end{table}

\begin{table*}[t]
\centering
\caption{Boosting strategy study results.}
\begin{tabular}{clcccccc}
\toprule
\textbf{} & \textbf{} & \textbf{CTR} & \textbf{PAY} & \textbf{GMV} & \textbf{Traffic Share} & \textbf{ROI} & \textbf{Hot Item Count} \\ 
\midrule
\multirow{2}{*}{\textbf{Rule}} 
    & w/o Exit  & -6.32\%  & -4.33\%  & -6.39\%  & -10.23\%  & -8.96\%  & -11.89\%  \\ 
    & w/o Promotion  & -4.21\%  & -5.12\%  & -4.46\%  & -11.22\%  & -13.12\% & -8.53\%   \\ 
\midrule
\multirow{4}{*}{\textbf{Number of Levels}} 
    & Fix Boosting             & -  & -  & -  & -   & -  &-   \\ 
    & 2  Stages           & +6.23\%  & +5.56\%  & +3.56\%  & +4.56\%   & +6.64\%  & +3.63\%   \\ 
    & 3 Stages                    & +9.22\%  & +10.26\% & +6.53\%  & +7.25\%   & +13.21\% & +8.96\%   \\ 
    & 4 Stages            & +9.28\%  & +10.36\% & +6.98\%  & +7.84\%   & +13.46\% & +8.78\%   \\ 
\bottomrule
\end{tabular}
\label{tab:rule_levels}
\end{table*}

\subsubsection{\textbf{Comparing Overall Platform Improvement}}
To validate the overall effectiveness gains of \model, we compared the performance of using AliBoost against a reverse experimental setting that did not adopt the AliBoost framework. By analyzing the overall statistical differences, we observed that cold items boosted by AliBoost over the past month contributed to platform-wide improvements, as illustrated in Table~\ref{tab:overall_ab}. Specifically, AliBoost achieved a 2.01\% increase in PV, a 4.51\% increase in CLICK, a 3.96\% increase in PAY, and a 4.69\% improvement in GMV. 

These results demonstrate that, on one hand, AliBoost successfully enhances cold items, while on the other hand, the boosted items are able to continuously earn effectiveness metrics like PV, CLICK, PAY, and GMV. This phenomenon shows that AliBoost provides substantial improvements to the ecosystem of the billion-scale Alibaba platform.

\subsubsection{\textbf{Comparing Cold Items Improvement}}
As shown in Table~\ref{tab:overall_ab}, AliBoost consistently makes natural recommendations across all metrics and time durations, with the improvement growing more significant as the cold items age.

Overall, AliBoost showed strong short-term impacts and sustained long-term benefits for cold items in the platform. For cold items launched within the first 3 days, AliBoost provided immediate boosts, achieving a +16.50\% increase in PV, +8.30\% in CLICK, +6.50\% in PAY, and +17.90\% in GMV. By 30 days, the improvements became more substantial, with +44.93\% in PV, +44.59\% in CLICK, +40.16\% in PAY, and +40.59\% in GMV.

In the mid-term, for items launched within 90 days, we observed continued growth, reaching +51.80\% in PV, +57.65\% in CLICK, +55.13\% in PAY, and +58.53\% in GMV. This demonstrates that AliBoost effectively enhances cold items even as they gain exposure over time.

Furthermore, in the long term, AliBoost sustained its superior performance for cold item boosting. For items launched within 180 days, the improvements peaked at +65.87\% in PV, +76.09\% in CLICK, +74.26\% in PAY, and +72.03\% in GMV, highlighting its ability to deliver long-lasting benefits.
These industrial results confirm that AliBoost not only addresses the cold-start problem but also drives significant engagement.

\subsubsection{\textbf{Freshness Improvement in the Alibaba Recommendation System}}
To further validate how \model improves the freshness of the Alibaba recommendation system, where freshness is defined as the traffic share of newly published items, we analyzed the proportion of traffic allocated to fresh items before and after the full implementation of \model in the system. 

\begin{figure}[tb]
    \centering
    \includegraphics[width=\linewidth, trim=0cm 0cm 0cm 0cm,clip]{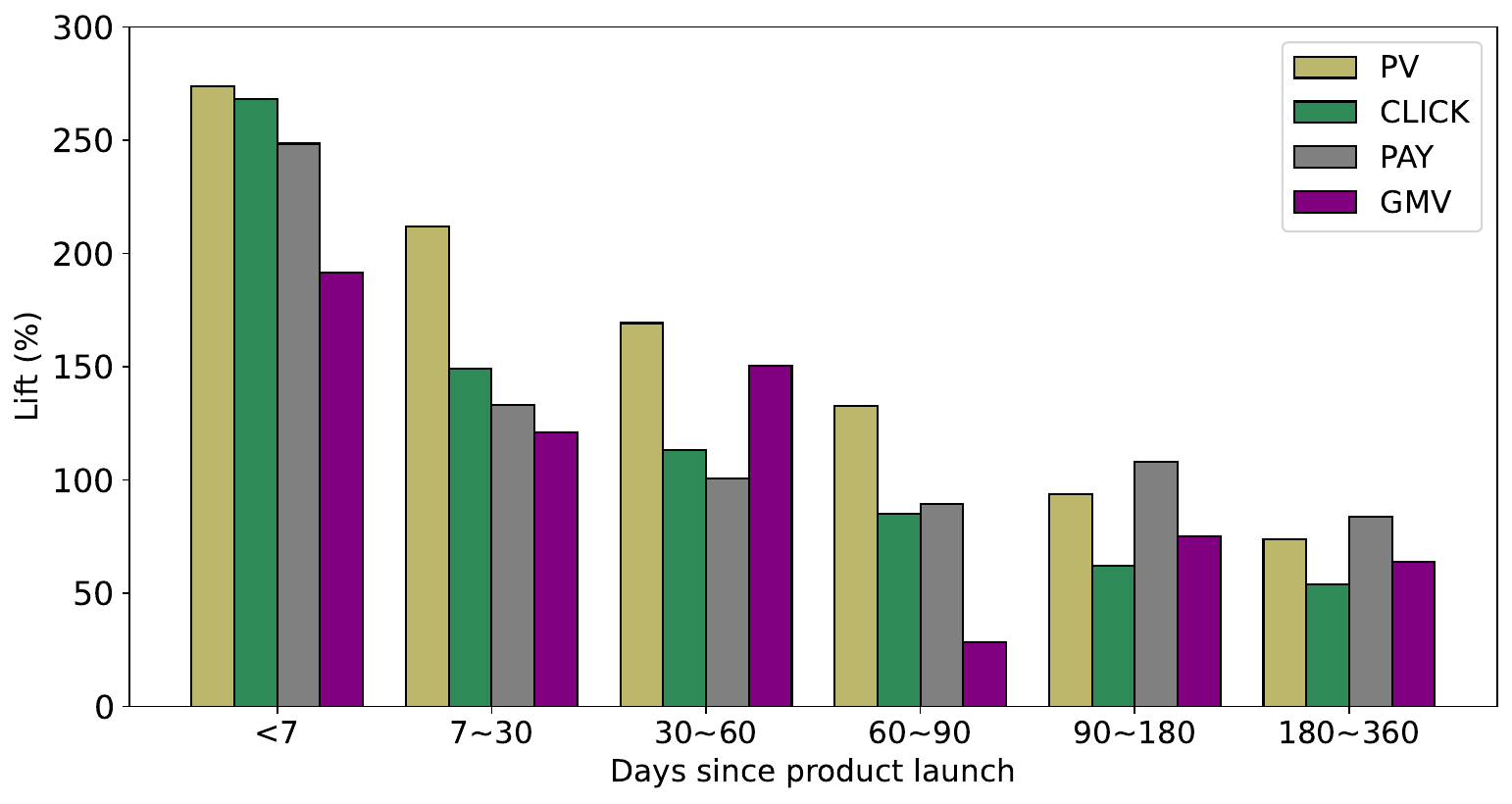}
    \caption{Relative changes in recommendation metrics for items with different release days on Taobao before and after the full-scale deployment of \model.}
    \label{fig:after_before_ab}
\end{figure}

As the results illustrated in~\autoref{fig:after_before_ab}, the exposure PV proportion of items launched within 7 days increased by over 250\%.
Moreover, the traffic share (exposure PV proportion) of items launched within 30 days increased by over 200\%. 
Similarly, significant increases were also observed in the Click, Pay, and GMV proportions.
These findings demonstrate that \model effectively improves the ecosystem of the recommendation system, significantly enhancing overall freshness for users.

\subsection{Effectiveness of Boosting Strategies (RQ2)}
In this subsection, we conduct an impact analysis to evaluate the effectiveness of our boosting strategies designed in \model, as illustrated in Table~\ref{tab:rule_levels}.

\label{sec:after_before_ab}

\subsubsection{\textbf{Impact of Promotion and Exit Rules}}
To evaluate the impact of the promotion and elimination rules on the overall performance of the \model framework, we conducted ablation experiments. The results indicate that removing either rule significantly affects the overall performance. Specifically, removing the elimination rule caused the CTR of cold-start items to decrease by over 6\%, ROI to decrease by over 8\%, and the number of hot items to decrease by over 11\%. Similarly, removing the promotion rule caused the CTR of cold-start items to decrease by over 4\%, ROI to decrease by over 13\%, and the number of hot items to decrease by over 8\%. These results demonstrate that both the promotion and elimination rules effectively enhance the efficiency of cold-start incubation.

\subsubsection{\textbf{Impact of Budget Segmentation Levels}}
We also conducted an experimental analysis on the effect of the number of budget segmentation levels in the cold-start exposure budget pool. The results show that as the number of segmentation levels increases, key efficiency metrics such as CTR, PAY, and ROI of cold-start items improve. Compared to no traffic segmentation (i.e., only 1 level), increasing the segmentation to 3 levels resulted in a 9\% improvement in CTR, a 10\% improvement in PAY, a 13\% improvement in ROI, and an increase of nearly 9\% in the number of hot items. However, due to the limited overall budget, increasing the number of segmentation levels beyond 3 showed diminishing returns. Therefore, we ultimately set the number of traffic segmentation levels to 3.

\begin{table}[tbp]
\centering
\caption{Performance improvement of stacking fine-tuning. Noted that the offline AUC is the same for two different time windows since the AUC is evaluated on the same set.}
\begin{tabular}{lccc}
\toprule
\textbf{Metric} & \textbf{Offline AUC} & \textbf{PV} & \textbf{PCTR} \\ 
\midrule
\textbf{3 Days Cold-Start (\%)} & +2.24\% & +8.30\% & +10.71\% \\ 
\textbf{7 Days Cold-Start (\%)} & +2.24\%  & +7.39\% & +11.40\% \\ 
\bottomrule
\end{tabular}
\label{tab:stf_result}
\end{table}

\subsection{Effectiveness of Stacking Fine-Tuning (RQ3)}
The results presented in Table~\ref{tab:stf_result} demonstrate the performance improvement offered by the stacking fine-tuning technique under different cold-start conditions over 3 and 7 days. The evaluation of our approach is based on the original basic online recommendation model (w/o stacking).

In terms of user engagement metrics, stacking fine-tuning achieves consistent positive gains in PV and PCTR. Specifically, for a 3-day cold-start period, PV improves by +8.30\%, while PCTR increases by an impressive +10.71\%, showcasing a strong capability to rapidly boost user interaction. Similarly, during the 7-day cold-start period, PV exhibits a +7.39\% improvement, and PCTR further strengthens with a +11.40\% increase, suggesting that the fine-tuning mechanism continues to enhance performance over extended time windows.
Regarding ranking quality, in an offline environment, we assessed the performance gains achieved by applying the stacking approach to the model currently deployed online.
The improvements in AUC suggest that the stacking fine-tuning effectively addresses both global and grouped ranking challenges. This positive impact is especially crucial in industrial recommendation systems where user behavior data can shift rapidly, and fine-tuning adaptability plays a critical role.

To more comprehensively assess the effectiveness of stacking, Appendix~\ref{sec:stack_baseline} presents offline experimental analyses built on representative CTR prediction and cold-start baselines.

\begin{table*}[t]
\centering
\caption{Ablation study on key factors of item-oriented bidding boosting.}
\begin{tabular}{lccccccc}
\toprule
\textbf{Exp Group} & \textbf{Rule} & \textbf{CTR} & \textbf{PAY} & \textbf{GMV} & \textbf{Traffic Share} & \textbf{ROI} & \textbf{Hot Item Count} \\ 
\midrule
1 & w/o Item-Oriented Bidding Boosting   & -45.31\% & -38.21\% & -42.11\% & -8.96\%  & -21.39\% & -9.36\%  \\ 
2& w/o Boosting Speed Factor        & -18.11\% & -14.98\% & -21.93\% & -4.09\%  & -8.09\%  & -4.67\%  \\ 
3& w/o User Preference Factor   & -28.38\% & -20.81\% & -26.03\% & -5.69\%  & -11.31\%  & -3.56\%  \\ 
\bottomrule
\end{tabular}
\label{tab:mechanism_impact}
\end{table*}

\subsection{Effectiveness of Item-Oriented Bidding Boosting (RQ4)}

The experimental results presented in Table~\ref{tab:mechanism_impact} highlight the critical contributions of individual system components to the performance of cold-start item incubation in terms of key metrics.

\subsubsection{\textbf{Impact of the Item-Oriented Bidding Boosting}} 
Removing the overall Item-Oriented Bidding Boosting mechanism resulted in the most severe performance degradation among the tested ablation settings. Specifically, CTR dropped by an extraordinary 45.31\%, while PAY and GMV decreased by 38.21\% and 42.11\%, respectively. Traffic Share also declined by 8.96\%, ROI by 21.39\%, and the Hot Item Count by 9.36\%. These findings underline the essential role of this bidding boosting mechanism, which strategically restricts item exposure based on flow value, ensuring optimized traffic allocation and effective budget utilization to maximize cold-start item visibility and growth.

\subsubsection{\textbf{Impact of the Boosting Speed Factor}}
Disabling the Boosting Speed Factor resulted in broad performance declines, though the effects were less pronounced compared to the removal of the Flow Value Threshold. CTR and GMV dropped by 18.11\% and 21.93\%, respectively, while ROI and Traffic Share were reduced by 8.09\% and 4.09\%. This suggests that regulating the speed of budget consumption is critical for maintaining a steady learning process, enabling the system to better match cold-start items to potential users over time and improving traffic efficiency.

\subsubsection{\textbf{Removing the User Preference Factor}} 
Eliminating the User Preference Factor module also led to significant performance deterioration, demonstrating the importance of identifying and prioritizing high-quality users. CTR fell by 28.38\%, with PAY, GMV, and ROI decreasing by 20.81\%, 26.03\%, and 11.31\%, respectively. Furthermore, the Hot Item Count dropped by 3.56\%, reflecting the diminished ability of the system to incubate promising items effectively. This module's ability to evaluate user activity and fatigue, and to dynamically adjust distribution thresholds, is essential for concentrating traffic on valuable, engaged users.

To further examine \model's boosting effects at a finer granularity across different product categories, Appendix~\ref{sec:boost_cate} presents a detailed category-level boosting case study.

\subsection{Mitigation of the Matthew Effect (RQ5)}
To validate how \model mitigates the Matthew Effect~\cite{gao2023alleviating,perc2014matthew}, the results summarized in Table~\ref{tab:aliboost_relative_changes} provide compelling evidence of how \model mitigates the Matthew Effect by reducing the dominance of top-exposed items over time. 

The results demonstrate that \model effectively mitigates the Matthew Effect by reducing the dominance of top-ranked items over time, thus enabling a more equitable distribution of exposure.
For items ranked in the Top 100 based on daily PV exposure, their retention rate drops significantly over longer observation periods. After 7 days, only 28.4\% of the Top 100 items remain, corresponding to a 71\% reduction. This trend continues over time, with retention rates further diminishing by 65.1\%, 53.8\%, and 42.7\% after 14, 21, and 30 days, respectively. Similarly, for items ranked in the Top 1000, the retention rates follow a comparable pattern, reducing by 56.3\% after 7 days and by 43.2\%, 36.9\%, and 29.2\% after 14, 21, and 30 evaluation days.

The consistent reduction in the dominance of top-ranked items demonstrates that \model disrupts the positive feedback loop typically associated with the Matthew Effect, where high-exposure items tend to monopolize future visibility. By actively redistributing exposure opportunities, \model ensures a more equitable distribution of traffic, thereby fostering diversity in the recommendation landscape. This mechanism proves particularly valuable in preventing over-concentration of exposure and promoting opportunities for high-quality but lower-ranked or cold-start items.

\begin{table}[t]
\centering
\caption{Relative decrease in the proportion of daily top-ranked items (based on PV exposure) that remain in the Top 100 and Top 1000 positions after 7, 14, 21, and 30 days, following the full deployment of \model.}
\begin{tabular}{lcccc}
\toprule
        & \textbf{7 Days} & \textbf{14 Days} & \textbf{21 Days} & \textbf{30 Days} \\ 
\midrule
\textbf{Top 100}  & -71.6\%   & -65.1\%   & -53.8\%   & -42.7\%   \\ 
\textbf{Top 1000} & -56.3\%   & -43.2\%   & -36.9\%   & -29.2\%   \\ 
\bottomrule
\end{tabular}
\label{tab:aliboost_relative_changes}
\end{table}

\section{Conclusion}
In this paper, we introduced AliBoost, an innovative ecological boosting framework designed to systematically enhance the visibility and growth of cold items on large-scale online platforms. By prioritizing exposure to promising cold items through a non-disturbing, performance-driven approach, AliBoost integrates potential prediction and user candidate selection mechanisms to effectively counteract the "rich-get-richer" effect.
The implementation of an Item-Oriented Bidding Boosting mechanism allows for a flexible and adaptive boosting process that accommodates the randomness of user behavior. Over the past six months, AliBoost has been deployed across Alibaba's mainstream platforms, successfully cold-starting over a billion new items and increasing both clicks and GMV of cold items by over 60\% within 180 days. Our extensive online A/B testing results demonstrate that AliBoost not only improves user experience and platform revenue but also contributes to a more balanced and flourishing ecosystem in Alibaba.

\begin{acks}
This work was supported in part by the National Natural Science Foundation of China (No. 62272200) and the Alibaba Taobao recommender systems.
\end{acks}

\bibliographystyle{ACM-Reference-Format}
\balance
\bibliography{sample-base}

\clearpage
\appendix
\section{Implementation Details in Alibaba}\label{sec:online}

In this subsection, we introduce the online implementation of our proposed \model framework, deployed in the homepage feed of Taobao (Alibaba's mainstream platform). The system architecture is designed to handle up to 120,000 QPS at traffic peaks, with an average response time of less than 20 milliseconds. This framework now serves as the backbone of Taobao's main traffic, addressing the cold-start problem for hundreds of millions of users and billions of items daily.

\begin{figure}[tb]
    \centering
    \includegraphics[width=\linewidth, trim=0cm 0cm 0cm 0cm,clip]{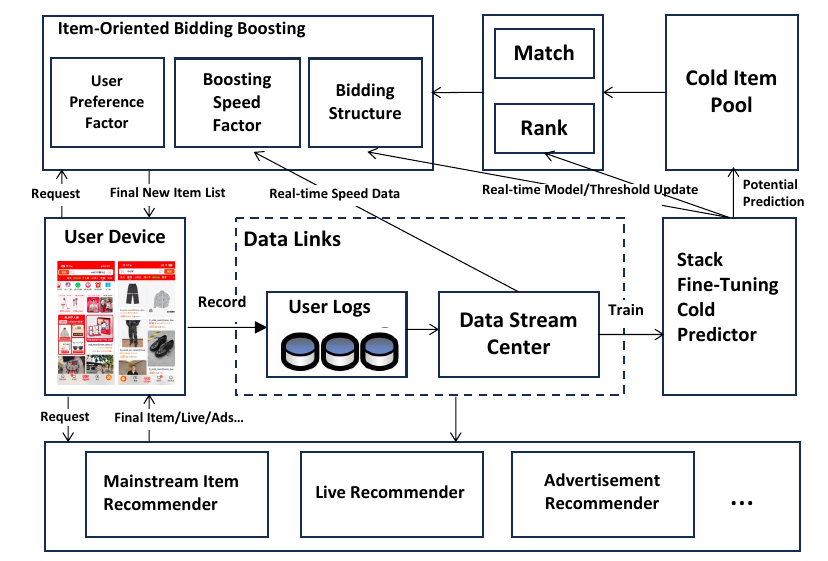}
    \caption{The deployment architecture of \model.}
    \label{fig:online_imple}
\end{figure}

The overall architecture of the Online RS is illustrated in Fig.~\ref{fig:online_imple}, which highlights its integration with various recommendation modules and real-time processing capabilities. The architecture consists of upstream recommendation subsystems, the Item-Oriented Bidding Boosting, Stack Fine-Tuning Cold Predictor, and supporting components for data processing and model updates.

\paragraph{Online Recommendation Subsystems}
Each time a user triggers a request, the upstream recommendation systems for different channels (e.g., cold item recommendation, mainstream item recommendation, short video recommendation, live recommendation, and advertisement recommendation) independently generate candidate items using their respective models. The results from each subsystem are sent to a unified ranking module for final sorting. Our \model framework is deployed specifically in the cold item recommendation channel to address the cold-start problem.
For the cold item channel, the process begins by retrieving a candidate pool of cold items relevant to the user's attributes and recent behaviors. These candidates are scored by a ranking model, which evaluates the match between each item's features and the user's preferences. The top-ranking candidates (typically a few thousand items) are then passed to the item-oriented bidding boosting framework for further evaluation and filtering.

\paragraph{Item-Oriented Bidding Boosting}
The Item-Oriented Bidding Boosting is the core module of the \model framework. For each candidate cold item, this module evaluates whether the item bidding price exceeds the user's ideal price. Specifically, the system compares the ranking model's score for each item with its corresponding threshold (basic score of ideal price). Items with scores below the threshold are discarded, while those exceeding the threshold are ranked, and the top 100 items are selected for final delivery to the unified ranking module.
The threshold for each item is dynamically adjusted in real time based on two key factors:
(1) Boosting speed factor: This mechanism ensures that each item's delivery speed aligns with its predefined target speed. If an item's delivery speed exceeds the target, the threshold is increased to reduce its exposure. Conversely, if the delivery speed falls below the desired level, the threshold is lowered to accelerate its exposure. Only in this way can the budget for the cold item avoid being exhausted in a very short time.
(3) User preference factor: To prioritize high-quality traffic, the system adjusts based on the user's activity level and fatigue. For highly active users with low fatigue, the threshold is reduced, allowing more items to be delivered. For less active or fatigued users, the threshold is increased to limit exposure.

\paragraph{Real-Time Data Processing and Model Updates}
The system leverages real-time data streams to continuously monitor and adapt its performance. Each user interaction is recorded in the user logs and sent to the Data Stream Center, which processes the data for both immediate updates and offline training. First, Stack Fine-Tuning CTR Model Updates: The Stack Fine-Tuning CTR Predictor, responsible for scoring cold items, is updated every few minutes using incremental training on the latest user interaction logs. These updates ensure that the model reflects the most recent trends and behaviors.
Second, Item Bidding Price Adjustment: The price values for all cold items are recalculated every 15 minutes based on the updated Stack CTR model and real-time delivery data. Additionally, the User Preference Factor mechanism computes the delivery speed for each item using a sliding window of 15 minutes, ensuring timely adjustments to prevent over- or under-delivery.

\paragraph{Scalability and Efficiency}
The Boosting Recommendation Framework is designed to operate efficiently at scale, seamlessly integrating with Taobao's billion-scale recommendation ecosystem. By focusing on cold items and leveraging the item-oriented bidding, the framework effectively reduces computational overhead while maintaining precise control over delivery rates. Compared to traditional systems, this approach ensures balanced exposure for cold items while prioritizing high-quality user traffic.

\paragraph{Impact}
Since its deployment, the Boosting Recommendation Framework has significantly improved the performance of Taobao's cold item recommendation system. Notably, the framework has increased the exposure of high-potential items while maintaining a high click-through rate (CTR). Furthermore, the Boosting Speed Factor mechanism has reduced over-delivery instances, ensuring a more balanced and efficient distribution of traffic among cold items.
By combining real-time evaluation, adaptive thresholding, and scalable architecture, the Boosting Recommendation Framework provides a robust solution to the cold-start problem, enabling efficient and effective recommendations in large-scale e-commerce environments.

\section{Additional Experimental Results}

\subsection{Offline Stacking Evaluation}\label{sec:stack_baseline}
Due to its dynamic-foundation-stacking structure, to compare performance in the offline setting, we used the industrial dataset, splitting 20\% of the items as cold items and the remaining 80\% as warm items. We trained the foundation models (DeepFM~\cite{guo2017deepfm} and DIN~\cite{zhou2018din}) using the data of the warm items. 
Further, we used the representative cold-start models ALDI~\cite{huang2023aldi} and ColdLLM~\cite{huang2024large} as baselines.
For basic DeepFM and DIN, we initialized the embeddings randomly as the initial embeddings.
For ALDI and ColdLLM, we used them as embedding initializer generators to compute the initial embeddings for cold items. These embeddings were fixed during inference. Stacking CTR used DeepFM or DIN, (DeepFM or DIN)+ALDI, or (DeepFM or DIN)+ColdLLM as foundation models, which remained fixed during the experiments.
The stacking component was dynamically updated during inference.

To evaluate the CTR prediction performance for cold items, we simulated online recommendations by testing CTR performance sequentially and divided the evaluation into the following two phases. (1) Phase I: Testing the CTR prediction performance for the first one to three interactions. (2) Phase II: Testing the CTR prediction performance for the first four to all interactions. The results are illustrated in Table~\ref{tab:stack}.
From the results, we observe that stacking CTR models shows great improvement over the foundation CTR models, especially as the number of interactions increases.

\begin{table}[tbp]
  \centering
  \caption{Offline stacking performance (AUC).}
  \resizebox{\linewidth}{!}{
    \begin{tabular}{c|cc}
    \toprule
    Model & Cold-Start Phase I & Cold-Start Phase II \\
    \midrule
    DeepFM & 0.5470 & 0.5408 \\
    DeepFM (Stack) & \textbf{0.5585} & \textbf{0.5674} \\
    DeepFM+ALDI & 0.5732 & 0.5667 \\
    DeepFM+ALDI (Stack) & \textbf{0.5910} & \textbf{0.5983} \\
    DeepFM+ColdLLM & 0.6123 & 0.6086 \\
    DeepFM+ColdLLM (Stack) & \textbf{0.6298} & \textbf{0.6342} \\
    \midrule
    DIN   & 0.5681 & 0.5728 \\
    DIN (Stack) & \textbf{0.5884} & \textbf{0.5947} \\
    DIN+ALDI & 0.5967 & 0.6015 \\
    DIN+ALDI (Stack) & \textbf{0.6033} & \textbf{0.6139} \\
    DIN+ColdLLM & 0.6196 & 0.6231 \\
    DIN+ColdLLM (Stack) & \textbf{0.6357} & \textbf{0.6438} \\
    \bottomrule
    \end{tabular}%
    }
  \label{tab:stack}%
\end{table}%

\subsection{Fine-Grained Boosting Effects Analysis}\label{sec:boost_cate}

To further evaluate this, we have compared the Boosted 7-Day PV and Boosted 30-Day PV of the main categories on our platform. Based on the results in Table~\ref{tab:cate}, we observed the following:

\begin{itemize}[leftmargin=*]
    \item Jewelry \& Accessories, Pets, Sports \& Outdoor, Food \& Fresh, and Auctions benefit the most from the boosting effects. These are categories that users may have an interest in and care about, but they might not be recommended to users without the boosting mechanism.
    \item 3C \& Digital experienced a decline in PV performance, with a 4\% and 8\% drop at 7 and 30 days, respectively. However, after analyzing the data, we found that although the PV for this category decreased, the PCTR increased by over 6.5\%. This indicates that our system is now recommending 3C \& Digital products more accurately. Previously, this category had higher PV but less accurate recommendations, leading to inefficient exposure.
    \item Customized Products and Toys \& Hobbies showed weak or negative growth at 7 days but exhibited positive growth at 30 days. The primary reason is that these two categories usually have lower PCTR as they are more specialized for different user segments. Our \model system quickly exits the boosting stage for these categories, allowing them a more comfortable growth environment. This ensures they are recommended more accurately and receive better-targeted exposure over time.
\end{itemize}

\begin{table}[htbp]
  \centering
  \caption{Boosting results on fine-grained item categories.}
    \begin{tabular}{cc|cc}
    \toprule
    \multicolumn{2}{c|}{Boosting Performance} & 7-Day PV & 30-Day PV \\
    \midrule
    \multicolumn{2}{c|}{ALL} & 31\%  & 45\% \\
    \midrule
    \multirow{18}[2]{*}{\rotatebox{90}{Representative Categories}} & Apparel \& Fashion & 17\%  & 28\% \\
          & Fast Moving Consumer Goods & 23\%  & 48\% \\
          & Sports \& Outdoor & 50\%  & 79\% \\
          & Food \& Fresh & 72\%  & 94\% \\
          & Toys \& Hobbies & -3\%  & 34\% \\
          & Jewelry \& Accessories & 63\%  & 123\% \\
          & Home Furnishing & -2\%  & 33\% \\
          & 3C \& Digital & -4\%  & -8\% \\
          & Automobiles & 49\%  & 111\% \\
          & Fresh Flowers \& Plants & 13\%  & 53\% \\
          & Pets  & 58\%  & 181\% \\
          & Industrial Products & 43\%  & 52\% \\
          & Health & 22\%  & 82\% \\
          & Small Appliances & 40\%  & 49\% \\
          & Commercial Agriculture & 43\%  & 115\% \\
          & Education & 28\%  & 53\% \\
          & Auctions & 68\%  & 206\% \\
          & Customized Products & -28\% & 49\% \\
    \bottomrule
    \end{tabular}%
  \label{tab:cate}%
\end{table}%

\end{document}